\documentclass[aps,prl,twocolumn,amsmath,amssymb,superscriptaddress,nofootinbib,showpacs]{revtex4}
\usepackage[english]{babel}
\usepackage{times}
\usepackage{latexsym}
\usepackage{graphics}
\usepackage{subfigure}
\usepackage{epsfig}
\usepackage{color}

\newcommand{\epsfboxmod}[1]{\epsfbox{#1.eps}}

\newcommand{\infig}[2]{\begin{center}
                                    \mbox{ \epsfxsize #1 \epsfboxmod{#2}}
                                      \vspace{-0.8cm}
                                    \end{center}}

\renewcommand{\paragraph}[1]{\vspace{0.2cm} \noindent {\it #1}:}
\renewcommand{\subparagraph}[1]{{\it #1 - }}

\newcommand{\ie}{{\it i.e. }}

\newcommand{\vect}[1]{\mathbf{#1}}

\newcommand{\sinc}{\textrm{sinc}}

\newcommand{\Vmin}{V_\textrm{min}}
\newcommand{\vmin}{v_\textrm{min}}
\newcommand{\Vr}{V_\textrm{R}}
\newcommand{\sigmar}{\sigma_\textrm{\tiny R}}
\newcommand{\alphar}{\alpha_\textrm{\tiny R}}

\newcommand{\tsigmar}{\widetilde{\sigma}_\textrm{\tiny R}}

\newcommand{\gdD}[1]{g}
\newcommand{\UdD}[1]{U}
\newcommand{\muoned}{{\mu'}}

\newcommand{\PR}{P}
\newcommand{\invPR}{P^{-1}}

\begin{document}

\title{
Ultracold Bose Gases in 1D Disorder:
From Lifshits Glass to Bose-Einstein Condensate
}

\author{P.~Lugan}
\author{D.~Cl\'ement}
\author{P.~Bouyer}
\author{A.~Aspect}
\affiliation{Laboratoire Charles Fabry de l'Institut d'Optique,
CNRS and Univ. Paris-Sud,
Campus Polytechnique,
RD 128,
F-91127 Palaiseau cedex, France}
\homepage{http://www.atomoptic.fr}
\author{M.~Lewenstein}
\affiliation{ICREA and ICFO-Institut de Ci\`encies Fot\`oniques,
Parc Mediterrani de la Tecnologia,
E-08860 Castelldefels (Barcelona), Spain}
\author{L.~Sanchez-Palencia}
\affiliation{Laboratoire Charles Fabry de l'Institut d'Optique,
CNRS and Univ. Paris-Sud,
Campus Polytechnique,
RD 128,
F-91127 Palaiseau cedex, France}

\date{\today}

\begin{abstract}
We study an ultracold Bose gas in the presence of 1D disorder for
repulsive inter-atomic interactions varying from zero to the
Thomas-Fermi regime. We show that for weak interactions the
Bose gas populates a finite number of localized single-particle
Lifshits states, while for strong interactions a
delocalized disordered Bose-Einstein condensate is formed. 
We discuss the schematic quantum-state diagram and derive the 
equations of state for various regimes.
\end{abstract}

\pacs{05.30.Jp,03.75.Hh,64.60.Cn,79.60.Ht}

\maketitle

%%%%%%%%%%%%%%%%%%%%%%%%%%%%%%%%%%%%%%%%%%%%%%%%%%%%%%%%%%%%%%%%%%%%%%
% \subsection{Introduction}
Disorder is present in nearly all condensed-matter
systems due to unavoidable defects of the sustaining media. 
It is known not only to impair quantum flows
but also to lead to spectacular effects such as
Anderson localization \cite{anderson1958,localization1d,gang4}. 
In contrast to condensed-matter systems, ultracold atomic gases 
can be realized in the presence of {\it controlled} disorder 
or quasi-disorder \cite{grynberg}, 
opening possibilities for investigations of localization
effects \cite{localizationcold1,localizationcold2,lsp2005,paul2005,lsp2007} 
(for review see Ref.~\cite{ahufinger2005}). 
The first experimental studies of localization in disordered interacting Bose 
gases have been reported
in Refs.~\cite{inguscio1,clement2005,inguscio2,clement2006,schulte2005,fallani2006}.

One of the most fundamental issues in this respect concerns the
interplay between localization and interactions in many-body
quantum systems at zero temperature.
Without interactions, a quantum gas in a random potential 
populates localized states
\cite{anderson1958}, either a single state (in the case of bosons),
or many (fermions). Weak repulsive interactions lead to delocalization
but strong interactions in confined geometries lead to
Wigner-Mott-like localization \cite{fisher1989}. 
Surprisingly, even for weakly interacting Bose gases, where the mean-field 
Hartree-Fock-Gross-Pitaevskii-Bogolyubov-de Gennes (HFGPBdG)
description is expected to be valid,
there exists 
no clear picture of the localization-delocalization scenario.
Numerical calculations using the Gross-Pitaevskii equation (GPE)
suggest that the Bose gas wave-function at low densities is 
a superposition of localized states \cite{schulte2005}. 
It is thus natural to seek
the true ground state 
in the form of generalized HFGPBdG states,
for which the Bose gas populates various low-energy single-atom states.
In the presence of disorder,
they correspond to so-called {\it Lifshits states} (LS) \cite{lifshits1988}.

%-----------------------------------------%
\begin{figure}[b!]
\begin{center}
\infig{20em}{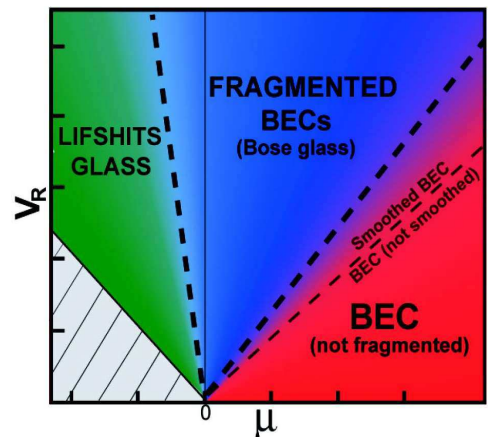}
\end{center}
\caption{(color online) Schematic quantum-state diagram of an interacting
ultracold Bose gas in 1D disorder. The dashed lines represent the
boundaries (corresponding to {\it crossovers}) 
which are controlled by the parameter 
$\alphar \!\! = \!\! \hbar^2/2m\sigmar^2 \Vr$
(fixed in the figure, see text),
where $\Vr$ and $\sigmar$ are the amplitude and correlation length
of the random potential.
The hatched part corresponds to a forbidden zone ($\mu \! < \! \Vmin$).
}
\label{fig:diagram}
\end{figure}
%-----------------------------------------%

In this Letter we 
consider a $d$-dimensional ($d$D)
Bose gas at zero temperature with repulsive interactions, and placed in a 1D
random potential with arbitrary
amplitude and correlation length.
We show that generalized HFGPBdG states indeed provide a very good description
of the many-body ground state for interactions varying from zero to
the Thomas-Fermi (TF) regime
\cite{noteTF}.
We stress that the solution we find is different from that of non-interacting
fermions which at zero temperature form a {\it Fermi glass} and occupy
a large number of localized single particle levels \cite{ahufinger2005}. 
In contrast, many bosons may occupy the same level
and thus populate only a finite number of LSs
forming what we call a {\it Lifshits glass}. 
In the following, we discuss the quantum states of the system as a function
of the strength of interactions and the amplitude and correlation length 
of the random potential,
and we draw the schematic quantum-state diagram (see
Fig.~\ref{fig:diagram}). In the limit of weak interactions,
the Bose gas is in the {\it Lifshits glass} state, whereas for
stronger interactions the gas forms a (possibly
smoothed) delocalized {\it disordered Bose-Einstein condensate} (BEC) \cite{lsp2006}. 
Our theoretical treatment provides us with a novel, physically clear, 
picture of disordered, weakly-interacting, ultracold Bose gases. 
This is the main result of this work.
In addition, we derive analytical formulae for the
boundaries (corresponding to crossovers) in the quantum-state diagram and 
for the equations of state in the various regimes. We illustrate our
results using a speckle random potential \cite{goodman}.

%%%%%%%%%%%%%%%%%%%%%%%%%%%%%%%%%%%%%%%%%%%%%%%%%%%%%%%%%%%%%%%%%%%%%%
% \subsection{Our system}
Consider a $d$D ultracold Bose gas with weak repulsive
interactions, \ie such as $n^{1-2/d} \! \ll \! \hbar^2\!/\!m\gdD{d}$, 
where $m$ is the atomic mass, $n$ the density
and $\gdD{d}$ the $d$D coupling constant.
The gas is assumed to be axially confined to a box of length
$2L$ in the presence of a 1D random potential $V(z)$,
and trapped radially in a 2D harmonic trap with frequency $\omega_\perp$. 
We assume that the random potential is
bounded below [$\Vmin \! = \! \min(V)$] and we use the scaling form 
$V(z) \! = \! \Vr v (z/\sigmar)$, where $v (u)$ is a random function with both 
typical amplitude and correlation length equal to unity \cite{noteparam}.

%%%%%%%%%%%%%%%%%%%%%%%%%%%%%%%%%%%%%%%%%%%%%%%%%%%%%%%%%%%%%%%%%%%%%%
% \subsection{The case of a speckle: an illustration}
For illustration, we will consider a 1D speckle potential 
\cite{goodman} similar to that used in 
Refs.~\cite{inguscio1,clement2005,inguscio2,clement2006}.
In brief, $v (u)$ is random with the probability
distribution 
$\mathcal{P}(v) \! = \! \Theta (v\!+\!1) \exp [ -(v\!+\!1) ]$,
where $\Theta$ is the Heaviside function.
Thus $v$ is bounded below by $\vmin\!\!=\!\!-\!1$ and we have
$\langle v \rangle\!\!=\!\!0$ and $\langle v^2 \rangle\!\!=\!\!1$.
In addition, for a square aperture, the correlation function reads
$\langle v(u) v(u') \rangle \! = \! \sinc^2\!\!\left[\!\sqrt{3/2}\left(u\!-\!u'\right)\!\right]$

Below, we discuss the quantum states of the Bose gas,
which are determined by the interplay of interactions and 
disorder.

%%%%%%%%%%%%%%%%%%%%%%%%%%%%%%%%%%%%%%%%%%%%%%%%%%%%%%%%%%%%%%%%%%%%%%
%\paragraph{Quantum phases of the Bose gas}

%%%%%%%%%%%%%%%%%%%%%%%%%%%%%%%%%%%
\subparagraph{BEC regime}
For strong repulsive interactions the Bose gas is delocalized and forms 
a BEC \cite{clement2005,lsp2006}
(possibly quasi-BEC in 1D or elongated geometries \cite{petrov2000s}).
The density profile is then governed by the GPE,
%+++++++++++++++++++++++++++++++++++++++++%
\begin{equation}
\mu = -\hbar^2\nabla^2(\!\sqrt{n})/2m\sqrt{n} + m\omega_\perp^2 \rho^2 / 2
+ V(\!z\!) + \gdD{d} n(\!\vect{r}\!),
\label{eq:GPEdensity}
\end{equation}
%+++++++++++++++++++++++++++++++++++++++++%
where $\rho$ is the radial coordinate and $\mu$ the chemical potential.
This regime has been studied
in the purely 1D case in Ref.~\cite{lsp2006}.
Here, we focus on the case of a shallow radial trap ($\hbar \omega_\perp \!\! \ll \!\! \mu$)
such that the radial profile is a TF inverted parabola.
Proceeding as in Ref.~\cite{lsp2006}, we find that the BEC density
has a generalized TF profile 
\cite{notemf1D}:
%+++++++++++++++++++++++++++++++++++++++++%
\begin{equation}
\sqrt{n(\rho,z)} \simeq
\sqrt{\mu (\rho)/\gdD{3}}\ [
1 - \widetilde{V} (\rho,z) / 2 \mu (\rho) ]
\label{eq:mfdD}
\end{equation}
%+++++++++++++++++++++++++++++++++++++++++%
where $\mu (\rho) \!\!\! = \!\!\! \mu \! \left[ 1 \!\! - \!\! (\rho/R_\perp)^2 \right]$ is
the local chemical potential,
$R_\perp \!\! = \!\! \sqrt{2\mu / m \omega_\perp^2}$ is the radial TF half-size
and $\widetilde{V} (\rho,z) \!\! = \!\! \int\! \textrm{d}z' G (\rho,z') V (z\!\!-\!\!z')$ 
is a {\it smoothed potential}
\cite{lsp2006} with 
$G(\rho,z) \!\! = \!\! \frac{1}{\sqrt{2} \xi (\rho)} \exp \left(\! -\frac{\sqrt{2}|z|}{\xi(\rho)} \!\right)$,
and 
$\xi(\rho) \!\! = \!\! \hbar / \sqrt{2m\mu (\rho)}$ 
being the local healing
length. 
For $\xi(\rho) \!\! \ll \!\! \sigmar$, \ie for
%+++++++++++++++++++++++++++++++++++++++++%
\begin{equation}
\mu (\rho) \gg {\hbar^2}/{2m\sigmar^2},
\label{eq:TFcond}
\end{equation}
%+++++++++++++++++++++++++++++++++++++++++%
we have
$\widetilde{V} (\rho,z) \! \simeq \! V (z)$,
and the BEC density follows the modulations of the random potential in the
TF regime. 
For $\xi(\rho) \!\! \gtrsim \!\! \sigmar$, the kinetic energy 
cannot be neglected
and competes with the disorder and the interactions. 
The
random potential is therefore smoothed \cite{lsp2006}: 
$\Delta \widetilde{V} (\rho) \!\! < \!\! \Delta V$
where $\Delta V$ ($\Delta \widetilde{V} (\rho)$) 
is the standard deviation of the (smoothed) random potential.
The solution~(\ref{eq:mfdD}) corresponds to a delocalized {\it
disordered BEC}.

The perturbative approach is valid 
% when the last term in Eq.~(\ref{eq:mfdD}) is much smaller than unity, \ie 
when $\mu (\rho) \!\! \gg \!\! \Delta \widetilde{V} (\rho)$.
From the expression for $\widetilde{V}$, we
write $\Delta \widetilde{V}(\rho) \!\! = \!\! \Vr \sqrt{\Sigma^0
(\sigmar / \xi (\rho))}$. 
For the speckle potential,
we can approximate the correlation function to
$\Vr^2\exp\left(-\!z^2/2\sigmar^2\right)$ and we find \cite{lsp2006}
%+++++++++++++++++++++++++++++++++++++++++%
\begin{equation}
\Sigma^0 (\tsigmar) = \tsigmar^2 + (1-2\tsigmar^2) \tsigmar \
\textrm{e}^{\tsigmar^2} \int_{\tsigmar}^\infty \textrm{d}\theta \
\textrm{e}^{-\theta^2}, \label{eq:mfdDcond2}
\end{equation}
%+++++++++++++++++++++++++++++++++++++++++%
with $\tsigmar \!\! = \!\! \sigmar / \xi (\rho)$. 
In the center, \ie $\rho \! = \! 0$ or in 1D, 
the validity condition of the BEC regime thus reduces to
%+++++++++++++++++++++++++++++++++++++++++%
\begin{equation}
\mu \gg \Vr \sqrt{\Sigma^0 (\sigmar/\xi)}
\textrm{~~~with~~~} \xi \! = \! \xi (0).
\label{eq:mfdDcond1}
\end{equation}
%+++++++++++++++++++++++++++++++++++++++++%

% \subparagraph{Fragmented regime}
If condition~(\ref{eq:mfdDcond1}) is not fulfilled,
the Bose gas will form a {\it fragmented BEC}. 
The latter is a compressible insulator and thus
can be identified with a {\it Bose glass} \cite{fisher1989}.

%%%%%%%%%%%%%%%%%%%%%%%%%%%%%%%%%%%
\subparagraph{Non-interacting regime}
In the opposite situation, for vanishing interactions, the problem
is separable and the radial wave-function is the ground
state of the radial harmonic oscillator. We are thus left with the eigenproblem 
of the single-particle 1D Hamiltonian
$\widehat{h} \!\! = \!\! -\hbar^2 \partial_z^2/2m + V(z)$.
In the presence of disorder, the eigenstates $\chi_\nu$ are 
all localized \cite{localization1d}
and are characterized by \cite{lifshits1988}
(i) a finite localization length,
(ii) a dense pure point density of states $\mathcal{D}_{2L}$,
and
(iii) a small participation length
$\PR_\nu \!\! = \!\! 1/\int \!\textrm{d}z \left| \chi_\nu (z) \right|^4$
\cite{notePR}.
If $V(z)$ is bounded below,
so is the spectrum 
and the low energy states belong to
the so-called {\it Lifshits tail},
which is 
characterized by a stretched exponential cumulative density of states (cDOS),
$\mathcal{N}_{2L} (\epsilon)
\!\! = \! \!\int_{}^{\epsilon}\! \textrm{d}\epsilon ' \mathcal{D}_{2L} (\epsilon ')
\! \sim \! \exp\!\left( \!-c \sqrt{\!\frac{\Vr}{\epsilon-\Vmin}} \right)$, in 1D
\cite{LSs}.

%-----------------------------------------%
\begin{figure}[t!]
\begin{center}
\infig{27em}{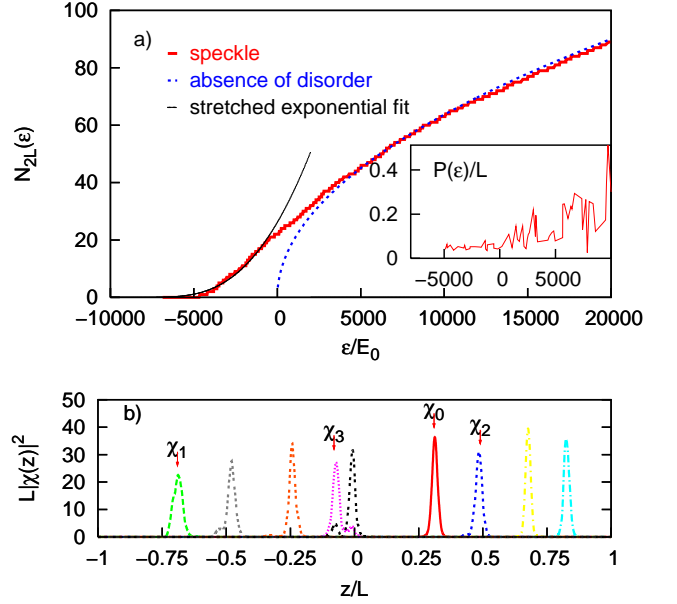}
\end{center}
\caption{(color online) 
a) Cumulative density of states of single particles in a speckle potential
with $\sigmar \!\! = \!\! 2 \! \times \! 10^{-3}L$ and $\Vr \!\! = \!\! 10^{4}E_0$,
where $E_0=\hbar^2/2mL^2$ ($\Vmin \!\! = \!\! -\Vr$). 
Inset: Participation length \cite{notePR}. 
b) Low-energy Lifshits eigenstates. 
For the considered realization of disorder, $\epsilon_0 \simeq -5\times \! 10^{3}E_0$.}
\label{fig:energies}
\end{figure}
%-----------------------------------------%

Numerical results for the speckle potential are shown in Fig.~\ref{fig:energies}. 
As expected the cDOS shows
a stretched exponential form, the lowest LSs are spatially localized, 
and $P(\epsilon)$
increases with energy indicating a weaker localization. 
However, $P(\epsilon)$ is almost constant at low
energy. Note also that the lowest LSs hardly overlap
if their extension is much smaller than the system size.

%%%%%%%%%%%%%%%%%%%%%%%%%%%%%%%%%%%
\subparagraph{Lifshits regime} 
We turn now to the regime of finite but weak interactions, 
where the chemical potential $\mu$ lies in the
Lifshits tail of the spectrum. Owing to the fact that 
the lowest single-particle LSs hardly overlap,
it is convenient to work in the basis of the LSs, $\{\chi_\nu, \nu \in \mathbb{N}\}$. 
These can be regarded as trapping micro-sites populated with $N_\nu$ bosons
in the quantum state $\phi_\nu (\vect{\rho}) \chi_\nu(z)$ where
the longitudinal motion is frozen to $\chi_\nu$ and 
$\phi_\nu$ accounts for the radial extension in the micro-site $\nu$.
Therefore, the many-body wave-function is the Fock state
%+++++++++++++++++++++++++++++++++++++++++%
\begin{eqnarray}
|\Psi \rangle & = & 
\prod_{\nu \geqslant 0} (N_\nu !)^{-1/2}(b_\nu^\dagger)^{N_\nu} | \textrm{vac} \rangle
\label{eq:fragBEC}
\end{eqnarray}
%+++++++++++++++++++++++++++++++++++++++++%
where $b_\nu^\dagger$ is the creation operator in the state $\phi_\nu (\vect{\rho}) \chi_\nu(z)$
\cite{mo-mf-thechnique}.
Each $\phi_\nu$ can be a transverse 2D BEC 
for $N_\nu \!\! \gg \!\! 1$. However, the quantum state~(\ref{eq:fragBEC}) 
does not correspond generally to a single 3D BEC
since it does not reduce to 
$(N !)^{-1/2}(b_0^\dagger)^{N}| \textrm{vac} \rangle$.
Rather, the Bose gas splits into several fragments whose longitudinal shapes
are those of the LSs, $\chi_\nu$, and are hardly affected by
the interactions.

The mean-field energy associated to the state~(\ref{eq:fragBEC}) reads
%+++++++++++++++++++++++++++++++++++++++++%
\begin{eqnarray}
E[\Psi] & = & \sum_\nu N_\nu \!\! \int \vect{\textrm{d}\rho} \
\phi_\nu^*
\left(
\frac{-\hbar^2 \nabla_\perp^2}{2m} + \frac{m\omega_\perp^2 \rho^2}{2} + \epsilon_\nu
\right)
\phi_\nu \nonumber \\
& + & \sum_\nu \frac{N_\nu^2}{2} \int \vect{\textrm{d}\rho} \
\UdD{3}_\nu |\phi_\nu|^4, \label{eq:GPE3D}
\end{eqnarray}
%+++++++++++++++++++++++++++++++++++++++++%
where $\UdD{3}_\nu \!\! = \!\! \gdD{d}\! \int\! \textrm{d}z |\chi_\nu(z)|^4 \!\! 
= \!\! \gdD{d} \invPR_\nu$
is the local interaction energy in the LS $\chi_\nu$.
Minimizing $E[\Psi]$ for a fixed number of atoms
($E[\Psi] \! - \! \mu \! \sum_\nu \! N_\nu \rightarrow \textrm{min}$),
we find the equation
%+++++++++++++++++++++++++++++++++++++++++%
\begin{equation}
(\mu \! - \! \epsilon_\nu) \phi_\nu =
\left[-\hbar^2 \nabla_\perp^2 / 2m +
       m\omega_\perp^2\rho^2 / 2 + N_\nu \UdD{3}_\nu |\phi_\nu|^2
\right]\phi_\nu.
\label{eq:GPE2D}
\end{equation}
%+++++++++++++++++++++++++++++++++++++++++%
Solving the 2D GPE~(\ref{eq:GPE2D}) for each micro-site $\nu$,
one finds
the atom numbers $N_\nu$
and the 
wave-functions $\phi_\nu$.
As $\mu$ increases, $|\phi_\nu|^2$ will
turn continuously from a Gaussian (for $\hbar \omega_\perp \!\! \gg \!\! \mu$) 
into an inverted parabola (for $\hbar \omega_\perp \!\! \ll \!\! \mu$).

To discuss the validity condition of the Lifshits regime, let us
call $\nu^\textrm{max}$ the index of the highest LS such that all
lower LSs hardly overlap. The Lifshits description requires
the chemical potential $\mu$ to be small enough so that the number
of populated LSs is smaller than $\nu^\textrm{max}$, \ie if
%+++++++++++++++++++++++++++++++++++++++++%
\begin{equation}
\mathcal{N}_{2L} (\mu) \leq \nu^\textrm{max}.
\label{eq:lifshitscond1}
\end{equation}
%+++++++++++++++++++++++++++++++++++++++++%
If condition~(\ref{eq:lifshitscond1}) is not fulfilled, several populated LSs will
overlap and the Bose gas will start to form a fragmented BEC. 
Each fragment will be a superposition of LSs, and its shape
will be modified by the interactions.

Although both $\mathcal{N}_{2L}$ and $\nu^\textrm{max}$ may have complex dependencies
versus $\Vr$, $\sigmar$ and the model of disorder,
general properties can be obtained using scaling arguments.
We rewrite the single-particle problem as
%+++++++++++++++++++++++++++++++++++++++++%
\begin{equation}
(\epsilon_\nu/\Vr) \varphi_\nu (u) =
- \alphar \partial_u^2 \varphi_\nu (u)
+ v(u) \varphi_\nu (u),
\label{eq:H0scaling}
\end{equation}
%+++++++++++++++++++++++++++++++++++++++++%
where $u \!\! = \!\! z/\sigmar$, $\varphi_\nu (u) \! = \! \sqrt{\sigmar} \chi_\nu (z)$
and $\alphar \!\! = \!\! \hbar^2/2m\sigmar^2\Vr$. Thus, all characteristics of
the spectrum depend only on the parameter $\alphar$
after renormalization of energies and lengths.
Scaling arguments show that in the Lifshits tail
%+++++++++++++++++++++++++++++++++++++++++%
\begin{equation}
\mathcal{N}_{2L} (\epsilon) \! = \! (L / \sigmar) \zeta (\alphar, \epsilon / \Vr)
~~~\textrm{and}~~~
\nu^\textrm{max} \! = \! (L / \sigmar) \eta (\alphar),
\label{eq:scaling}
\end{equation}
%+++++++++++++++++++++++++++++++++++++++++%
where $\zeta$ and $\eta$ are $v$-dependent functions. Finally, inserting these expressions
into Eq.~(\ref{eq:lifshitscond1}) and solving formally, we obtain the
validity condition of the Lifshits regime:
%+++++++++++++++++++++++++++++++++++++++++%
\begin{equation}
\mu \leq \Vr F (\alphar), \label{eq:lifshitscond2}
\end{equation}
%+++++++++++++++++++++++++++++++++++++++++%
where $F$ is the solution of
$\zeta\big(\alphar,F(\!\alphar\!)\big) \!\! = \!\! \eta (\alphar)$, which can be
computed numerically, for example.

%%%%%%%%%%%%%%%%%%%%%%%%%%%%%%%%%%%%%%%%%%%%%%%%%%%%%%%%%%%%%%%%%%%%%%
% \subsection{Schematic phase diagram}
%\paragraph{Schematic phase diagram}
We are now able
to draw the schematic quantum-state
diagram of the zero-temperature Bose gas as a function of
$\mu$ and $\Vr$ (see Fig.~\ref{fig:diagram}). 
From the discussion above, it is clearly fruitful to fix the parameter
$\alphar$ while varying $\Vr$.
The boundaries between the various regimes 
(Lifshits, fragmented BEC, BEC and smoothed BEC)
result from the competition between the interactions and the disorder
and are given by Eqs.~(\ref{eq:TFcond},\ref{eq:mfdDcond1},\ref{eq:lifshitscond2}). 
We stress that they are crossovers rather than phase transitions.
Interestingly, all these boundaries are straight
lines with slopes depending on the parameter $\alphar$. 
This is
clear from Eq.~(\ref{eq:lifshitscond2}) for the boundary between the
Lifshits and the fragmented regimes. In addition, since 
$\Vr \!\! = \!\! (\hbar^2/2m\sigmar^2)/\alphar$, the non-smoothing
condition~(\ref{eq:TFcond}) reduces to $\mu \! \gg \! \alphar \Vr$.
Finally, since $\mu \! = \! \hbar^2/2m\xi^2$ and thus 
$\sigmar/\xi = \frac{1}{\sqrt{\alphar}} \sqrt{\mu / \Vr}$, 
the non-fragmented BEC
condition~(\ref{eq:mfdDcond1}) also corresponds to a straight line
with a slope depending on $\alphar$ in Fig.~\ref{fig:diagram}.

%%%%%%%%%%%%%%%%%%%%%%%%%%%%%%%%%%%%%%%%%%%%%%%%%%%%%%%%%%%%%%%%%%%%%%
% \subsection{Equations of state}
To finish with, we derive the equations of state of the Bose gas in the 
identified quantum states. 
It is important to relate the chemical potential $\mu$ which governs
the crossovers between the
various regimes to the mean atomic
density $\overline{n}=N/2L$ and the coupling constant $g$. 
Both can be controlled in experiments with ultracold atoms.

%%%%%%%%%%%%%%%%%%%%%%%%%%%%%%%%%%%
\subparagraph{Tight radial confinement} 
For
$\muoned \!\! - \!\! \epsilon_\nu \!\! \ll \!\! \hbar \omega_\perp$ 
where 
$\muoned \!\! = \!\! \mu \!\! - \!\! \hbar\omega_\perp$, 
the
radial wave-functions are frozen to zero-point
oscillations, $\phi_\nu (\vect{\rho})=\exp(-\rho^2/2
l_\perp^2)/\sqrt{\pi} l_\perp$ with
$l_\perp \!\! = \!\! \sqrt{\hbar/m\omega_\perp}$ the width of 
the radial oscillator.

In the BEC regime, $\muoned \gg \Delta\widetilde{V}$, we find from
Eq.~(\ref{eq:mfdD}),
%+++++++++++++++++++++++++++++++++++++++++%
\begin{equation}
\muoned = \overline{n}\gdD{1}.
\label{eq:mu1Dcontspeckle1}
\end{equation}
%+++++++++++++++++++++++++++++++++++++++++%

In the Lifshits regime, we find
%+++++++++++++++++++++++++++++++++++++++++%
\begin{equation}
N_\nu \!\! = \!\! [\muoned \! - \! \epsilon_\nu]/\UdD{1}_\nu
\textrm{~for~} \muoned>\epsilon_\nu 
\textrm{~~~and~~~}
N_\nu \! = \! 0  \textrm{~otherwise},
\label{eq:alpha1D}
\end{equation}
%+++++++++++++++++++++++++++++++++++++++++%
by inserting the above expression for $\phi_\nu(\rho)$ into Eq.~(\ref{eq:GPE2D}).
Turning to a continuous formulation and using the normalization condition,
$N \!\! = \!\! \int\! \textrm{d}\epsilon \mathcal{D}_{2L}(\epsilon) N(\epsilon)$, we deduce
the equation of state of the Bose gas in the Lifshits regime:
%+++++++++++++++++++++++++++++++++++++++++%
\begin{equation}
N \gdD{1} = \int_{-\infty}^{\muoned} \textrm{d}\epsilon\
\mathcal{D}_{2L}(\epsilon) \left( \muoned - \epsilon \right) P(\epsilon),
\label{eq:mu1Dcont}
\end{equation}
%+++++++++++++++++++++++++++++++++++++++++%
which relates the chemical potential
$\muoned$ to the coupling constant $\gdD{1}$.
The relation
is in general non-universal (\ie it depends on the model of disorder, $v$).
In the case of a speckle potential, 
$\mathcal{N}_{2L}(\epsilon) \!\! = \!\! A(\alphar)\! (L / \sigmar)
\exp\left( \!-c(\alphar) / \sqrt{\epsilon/\Vr \!\! + \!\! 1}\right)$,
and assuming that the participation length 
$\PR_\nu \!\! = \!\! \sigmar p_\nu (\alphar)$ is
independent of the energy in the Lifshits tail, we find:
%+++++++++++++++++++++++++++++++++++++++++%
\begin{equation}
\overline{n} \gdD{1} \simeq A(\!\alphar\!) c^2(\!\alphar\!) p_0(\!\alphar\!) \Vr \
\Gamma\left( -2, c(\alphar) / \sqrt{\mu'/\Vr +1} \right),
\label{eq:mu1Dcontspeckle2}
\end{equation}
%+++++++++++++++++++++++++++++++++++++++++%
where $\Gamma$ is the incomplete gamma function
and 
$A(\alphar)$, $c(\alphar)$ and $p_0(\alphar)$ can determined
numerically.

\begin{figure}[t!]
\begin{center}
\infig{27em}{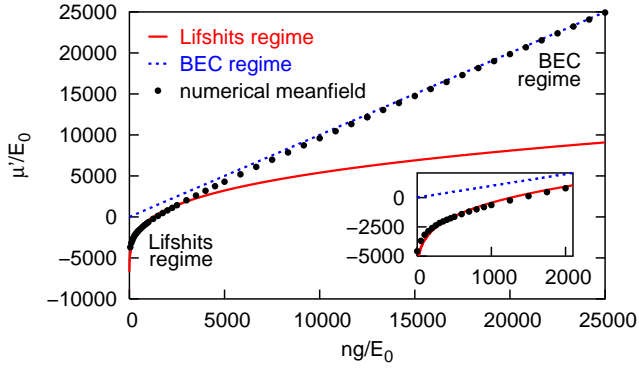}
\end{center}
\caption{(color online)
Chemical potential of a Bose gas in a speckle potential with the same parameters as in Fig.~\ref{fig:energies}, in the case of a tight radial confinement
($\mu' \! - \! \epsilon_\nu \! \ll \! \hbar \omega_\perp$).
The points are given by numerical calculations; the solid and dotted lines
represent the analytical formulae Eqs.~(\ref{eq:mu1Dcontspeckle2}) and 
(\ref{eq:mu1Dcontspeckle1}) derived for the Lifshits and BEC regime, respectively.
}
\label{fig:muVSg1D}
\end{figure}

Using a numerical minimization of the energy functional~(\ref{eq:GPE3D}) in the
Gross-Pitaevskii formulation, we compute the
chemical potential of the Bose gas in a wide range of interactions.
The result shown in Fig.~\ref{fig:muVSg1D} indicates a
clear {\it crossover} from the Lifshits regime to the BEC
regime as the interaction strength increases. The
numerically obtained chemical potential $\mu$ agrees
with our analytical formulae in both Lifshits and BEC regimes.

%-----------------------------------------%
%-----------------------------------------%

%%%%%%%%%%%%%%%%%
\subparagraph{Shallow radial confinement} 
The equations of state
can also be obtained in the case of shallow
radial confinement ($\mu - \epsilon_\nu \!\! \gg \!\! \hbar \omega_\perp$). In the BEC regime, for $\xi \!\! \ll \!\! \sigmar$, we
find $\mu \!\! \simeq \!\! \sqrt{\overline{n} \gdD{3}
m\omega_\perp^2/\pi \!\! - \!\! \Vr^2}$. In the Lifshits regime the 2D
wave-functions $\phi_\nu(\vect{\rho})$ are in the TF regime,
$|\phi_\nu(\vect{\rho})|^2 \!\! = \!\! \frac{\mu-\epsilon_\nu}{N_\nu
\UdD{3}_\nu}\left(1\!-\!\rho^2/R_\nu^2\right)$,
where $R_\nu \!\! = \!\! \sqrt{2(\mu \! - \! \epsilon_\nu)/m\omega_\perp^2}$ is the 2D-TF radius
and $N_\nu \! = \! \pi (\mu \! - \! \epsilon_\nu)^2 / m\omega_\perp^2 \UdD{3}_\nu$
for $\mu \! > \! \epsilon_\nu$ ($0$ otherwise).
Proceeding as in the 1D case, we find:
%+++++++++++++++++++++++++++++++++++++++++%
\begin{equation}
N \gdD{3} = \frac{\pi}{m\omega_\perp^2}
\int_{-\infty}^{\mu} \textrm{d}\epsilon\
\mathcal{D}_{2L}(\epsilon) \left( \mu - \epsilon \right)^2
P(\epsilon).
\label{eq:mu3Dcont}
\end{equation}
%+++++++++++++++++++++++++++++++++++++++++%
Applying this formula to the relevant model of disorder allows us to compute
the populations $N_\nu$ of the various LSs $\chi_\nu$ and the corresponding 
radial extensions $\phi_\nu$.

In summary, we have presented a complete picture of the quantum states of an
interacting Bose gas
in the presence of 1D disorder, including the novel description of the
weakly interacting {\it Lifshits glass} state.
We have provided analytical formulae
for the boundaries (crossovers) in the quantum-state diagram 
and shown that they are determined by the coupling constant.
Since this coupling constant can be controlled in cold gases, 
future experiments should be able to explore the whole diagram.

%%%%%%%%%%%%%%%%%%%%%%%%%%%%%%%%%%%%%%%%%%%%%%%%%%
% \vspace{1.cm}
We thank G.~Shlyapnikov, L.~Santos, D.~Gangardt, C.~Mora,
R.~Hulet and R.~Nyman for useful discussions.
This work was supported by
the French DGA, MENRT, IFRAF and ANR,
the Spanish MEC Programs, DUQUAG, QOIT and 
the European Union QUDEDIS and IP SCALA.

%%%%%%%%%%%%%%%%%%%%%%%%%%%%%%%%%%%%%%%%%%%%%%%%%%%%%%%%%%%%%%%%%%%%%%

%%%%%%%%%%%%%%%%%%%%%%%%%%%%%%%%%%%%%%

%%%%%%%%%%%%%%%%%%%%%%%%%%%%%%%%%%%%%%%%%%%%%%%%%%%%%%%%%%%%%%%%%%%%


\begin{references}
\bibitem{noteparam}
The random potential $V(z)$ is thus
characterized by a typical amplitude $\Vr$ and a typical
correlation length $\sigmar$.
\bibitem{notePR}
The participation length scales as the
width of the wave-function and thus
decreases for stronger localization.
\end{references}
\end{document}